\documentclass[11pt]{article}
\usepackage{graphicx}
\usepackage{amsmath}
\usepackage{amssymb}
\usepackage{amsxtra}

\usepackage{adjustbox}

\title{What gives a ``theory of Initial Conditions'' ? }

\author{Holger Bech Nielsen\footnote{Speaker at the workshop 
``What comes beyond the Standard Models'' in Bled 2022.}\\
Niels Bohr Institute, University of Copenhagen \\
Blegdamsvej 17, Copenhagen $\O$, Denmark \\
and \\
  Keiichi Nagao\\
  Faculty of Education, Ibaraki University \\
Bunkyo 2-1-1, Mito 310-8512 Japan }

\date{``Bled''   , July , 2022}

\begin{document}
\maketitle

\begin{abstract}
  The present work contains a review of some of the work we have done
  on complex action or non-Hermitian Hamiltonian theory, especially
  the result that the anti-Hermitian part of the Hamiltonian functions
  by determining the actual solution to the equations of motion, that
  should be realized; this means it predicts the initial conditions.
  It should be stressed that a major result of ours is that the effective
  equations of motion will in practice - after long time - be so
  accurately as if we had indeed a Hermitian Hamiltonian, and so
  there is at first nothing wrong in assuming a non-Hermitian one. In fact
  it would practically seem Hermitian anyway. 
  A major new point is that we seek by a bit intuitively arguing to suggest
  some cosmologically predictions from the mentioned initial conditions
  predicted: We seek even by assuming essentially nothing but very general
  properties of the non-Hermitian Hamiltonian that we in practice should find a
  bottom in the (effective Hermitian) Hamiltonian and that the Universe
  at some moment should pass through a (multiple) saddle point very closely,
  so that the time spent at it would be very long.
\end{abstract}



\section{Introduction
}\label{s:intro}

It would be very nice to unify our knowledge of the equations of
motion, or we could say the time-development, with our knowledge about the
initial
conditions, or as we shall look upon it here, which solutions to the equations
of motion is by the initial-condition-physics selected as the one to be
realized, the true development.
We\cite{9,11,12,13,14,15,16,17,20,21,27,28,30,31,32,Nagao:2022rap,Nagao:2022jam} 
and also Masao Ninomiya\cite{1,2,3,4,5,6,7,8,10} have long
worked on the idea that the action should not be real, but rather complex.
It has turned out that such theories of complex action or essentially
similarly of non-Hermitian \footnote{The Hamiltonian is 
not restricted to the class of PT-symmetric non-Hermitian Hamiltonians that were 
studied in 
Refs.\cite{22,23,24,25,26}.} 
in fact lead to a theory in
which
\begin{itemize}
\item The effect of the non-hermiticity is not seen after appreciable time
  in the equations of motion, so that effective hermiticity basically came out
  automatically; and thus this kind of theory is indeed viable!
  \item But the initial conditions is predicted from such theories.
  \end{itemize}

But it is then of course very important for whether such a hypothesis of
of a complex action or equivalently non-Hermitian Hamiltonian can be
upheld, whether the action or the Hamiltonian can be arranged in a reasonable
way so as to give some initial condition informations matching with
what we know about the initial conditions having governed the
world, the universe.

It is the purpose of the present article in addition to reviewing our
works on this complex action type of theory to argue even without
making any true fitting of the Hamiltonian, except assuming it to have
a classical analogue - using in fact a phase space consideration - but
rather looking only at an essentially random form of the Hamiltonians,
especially the anti-Hermitian part, a not so bad crude picture of the initial
condition pops out. In fact what we call this crude success is that the
favored or likely initial arrangement becomes that the system - the world -
shall pass through and stay {\em very long} in saddle points. We namely
interpret this prediction as being optimistically the prediction
of the world going through an in some sense long stage of the inflation
situation. An inflaton field having the value equal to the maximum 
of the (effective) potential for the inflaton field represents namely
for each Fourier component of this inflaton  field a system sitting at
one of its saddle points. So indeed in the phenomenological development
of the universe it goes through a state, which is precisely a saddle point,
with respect to an infinity of degrees of freedom, namely the various Fourier
components. The fact that we predict a very slow going through might be
taken as an encouragement by comparison with, that it is a well-known
problem, ``the slow roll problem'', that the inflation for phenomenological
reasons should be kept going longer than expected unless the inflaton
effective potential is especially (and somewhat in the models constructed)
flat. Flatness should help to make the inflation period longer than it
would be `` naturally''. Our long staying prediction might be taken as one of 
benefits of our model with the complex action seeking to get a long inflation, even
with a less flat effective potential. 

In the following section \ref{initial} we shall talk about initial
conditions and give very crude arguments for a long staying saddle point
being favored.
In section \ref{intuitive} we draw some crude phase space configurations
from which we seek to get an idea about which behavior of the development
of the mechanical system with the complex action would be to expect.
We end with favoring  the long stay at the saddle point and going also to a
region near a (local) minimum in the effective Hamiltonian, thereby explaining
an effective finding of a bottom of the Hamiltonian.
Then, in section \ref{main}, we allude to our for the belief in our complex
action having a chance to be really true most important derivation: that you
would not observe any effect (except from the initial conditions) in practice
after sufficient time.
In section \ref{WV} we introduce the idea that when one includes the future
as one should in our model one may write an expression
for the expectation value of a dynamical variable as the by Aharonov et.al. 
introduced weak value\cite{18,19}. This is a possible scheme for extracting the
to be expected average for  experiment. A priori this kind of
weak values are complex, but we have made theorems which prove reality
under some assumptions.

\section{Hope of Making Theory of Initial Conditions}
\label{initial}

    The laws of physics falls basically in the two classes:
    \begin{itemize}
    \item The equations of motions including the possible states of the
      universe and thus the types of particles existing.
      (it is here we find the Standard Model).
    \item Laws about the initial conditions. Here we may think of the second
      law of thermodynamics, and perhaps some cosmological laws as the
      Hubble expansion. Or may be inflation.
      \end{itemize}

  We have long worked on the hypothesis that the Hamiltonian were not
  Hermitian. At first one thinks that this would have been seen immediately,
  but a major result of ours were:

  {\bf For the equations of motion there would be no clean signature of the
    Hamiltonian not being Hermitian left.}
  The only significant revelation of the imaginary part of the
  Hamiltonian would be {\bf via  the initial conditions}.
\vspace{1mm}

\noindent
  This then means that by such a non-Hermitian Hamiltonian theory we potentially
  has found a theory, that could function as a theory behind the initial
  state laws, we have at present.

\section{Intuitive Understanding}
\label{intuitive}

Let us give the reader an idea about what we have in mind by thinking of
a skier with frictionless skies, so that he can only stop when he has run up
the hill and lost the kinetic energy. Then there is some given distribution
of the quality of the outlook he can enjoy in different places or of some other
sort of attraction which the skier would like to enjoy as long as possible.

It is not a good idea to just start at a random attractive outlook point 
with splendid outlook, because the skier will most likely find that on the hill side and he
will rush down and thus away from the attractive outlook  point quickly. Even
arranging to slide first up with speed as to just stop by loosing the kinetic
energy at an attractive point, might not compete with finding an even a bit
less attractive outlook  but still very attractive outlook point at
the ``middle'' of a
pass, in which one can stand seemingly forever.Just a little accidental
slide to one side or the other of course in the pass leads to that he slides
down and the attractive outlook place soon gets lost. With quantum mechanics
such a slight leaving the very metastable saddle point in the pass is
unavoidable. So with quantum mechanics the skier has to plan that he cannot be
in the saddle point forever, but will slide out some day. Then he has to plan
for the next step what is most profitable. Presumably it is best to arrange
to find a reasonable attractive outlook place in a little whole in the
landscape
of only very little less potential energy than the starting situation chosen.
Then namely it could be arranged that the skier would only ski little and slowly
around when first arriving there. Presumably it would be best to then
if possible have arranged to get back again to the first saddle point in the pass
and perhaps cyclically repeat again and again a good trip.
\begin{figure}
  \includegraphics{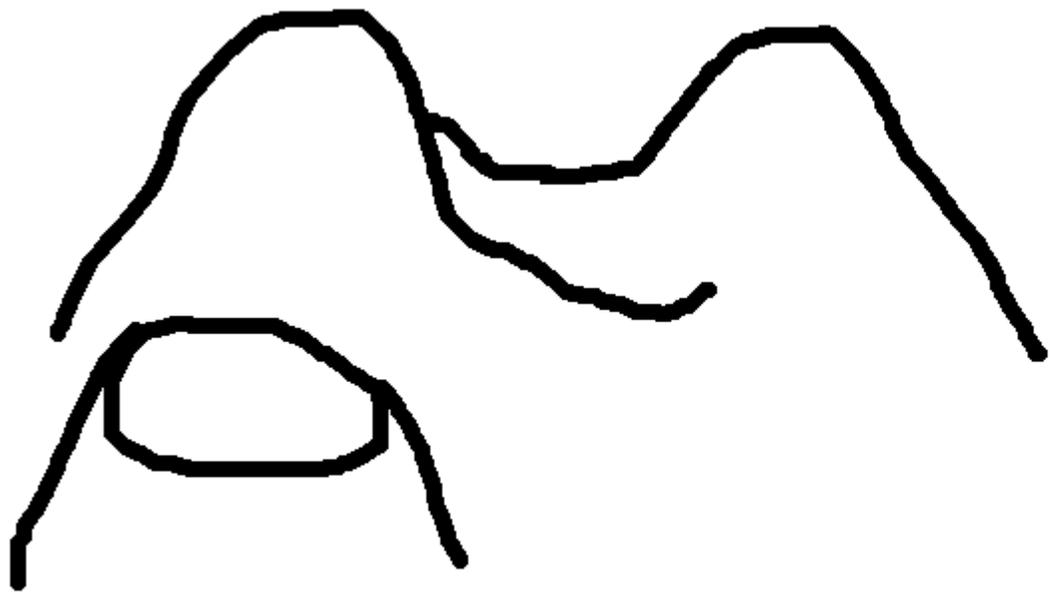}
  \caption{This just a skiing terrain, that should really symbolize in
    our work any state of the whole Universe. We imagine a little skier with
    frictionless skies, which can ski around but his tour is fixed from where and with what velocity he
    starts. He cannot stop except if he just runs up the hill and runs
    out of kinetic energy. }
\end{figure}

\begin{figure}
  \includegraphics{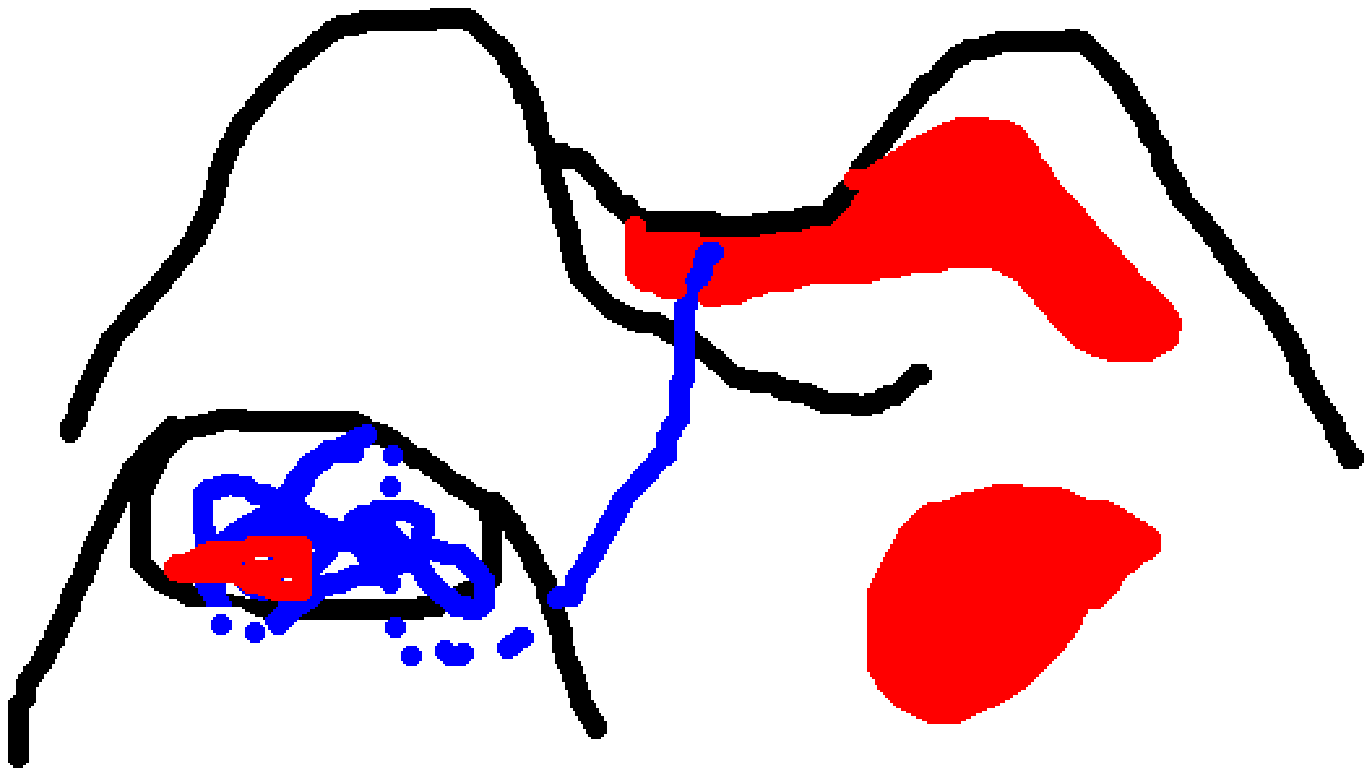}
  \caption{Now we have put on some red spots which are the regions with
    the best outlooks or for other reasons the best ones to stay in. Now the
    skier gets the problem of starting in such a clever way that he manages
    to stay the longest time in these the best regions (marked by red). Going
    just to a good region at random would probably mean that he very fast
    would rush out of it and it would only be a short enjoy of the good region.
    What to do?}
\end{figure}

\subsection{Phase space drawings}

We now present a few figures supposed to be drawn in phase space, rather
than in a geometrical space with mountains, to illustrate again the
considerations which the little skier has to do to get the most glorious
outlook for so long as possible. One must think of an integral over time
of a quantity measuring the beauty of the outlook now in different ``places''
in phase space because the outlook beauty degree can of course also depend
on the momentum, or say the velocity. In our theory with non-Hermitian
or in classical thinking complex Hamiltonian the quantity to be identified
with the beauty degree for the skier is of course the imaginary part $Im H $
of the Hamiltonian. This imaginary part $Im H$ namely enhances the
normalization of the wave function describing the the skier or in our
model say the universe as it moves along classically in the phase space.

Thus the route through phase space which maximizes the time integral
over the imaginary part $\int Im H dt$ is the one that makes the wave function
grow the most. This means that the chance for surviving the tour by the skier
or rather the universe the development of which is described by the tour has
the largest amplitude for existing at all at the end of the tour, when the
integral over time $ \int Im Hdt$ is the largest. It is therefore
our theory predicts that what really shall happen most likely is the
route through phase space giving this integral the maximal value.
So we see that a problem like the one for the little skier is set up.

\begin{figure}  
  \includegraphics[scale=0.7]{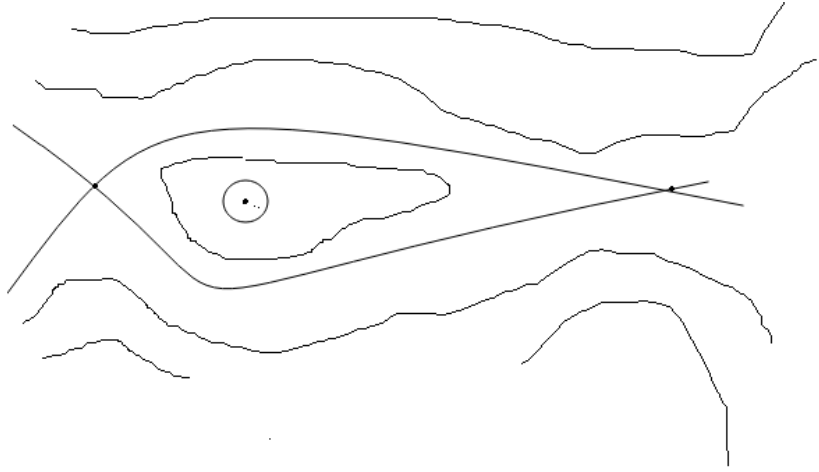}
  \caption{{\label{f3}\bf Symbolic Phase Space for Universe, Level curves for ReH}}
\end{figure}

\begin{figure}
\includegraphics[scale=0.7]{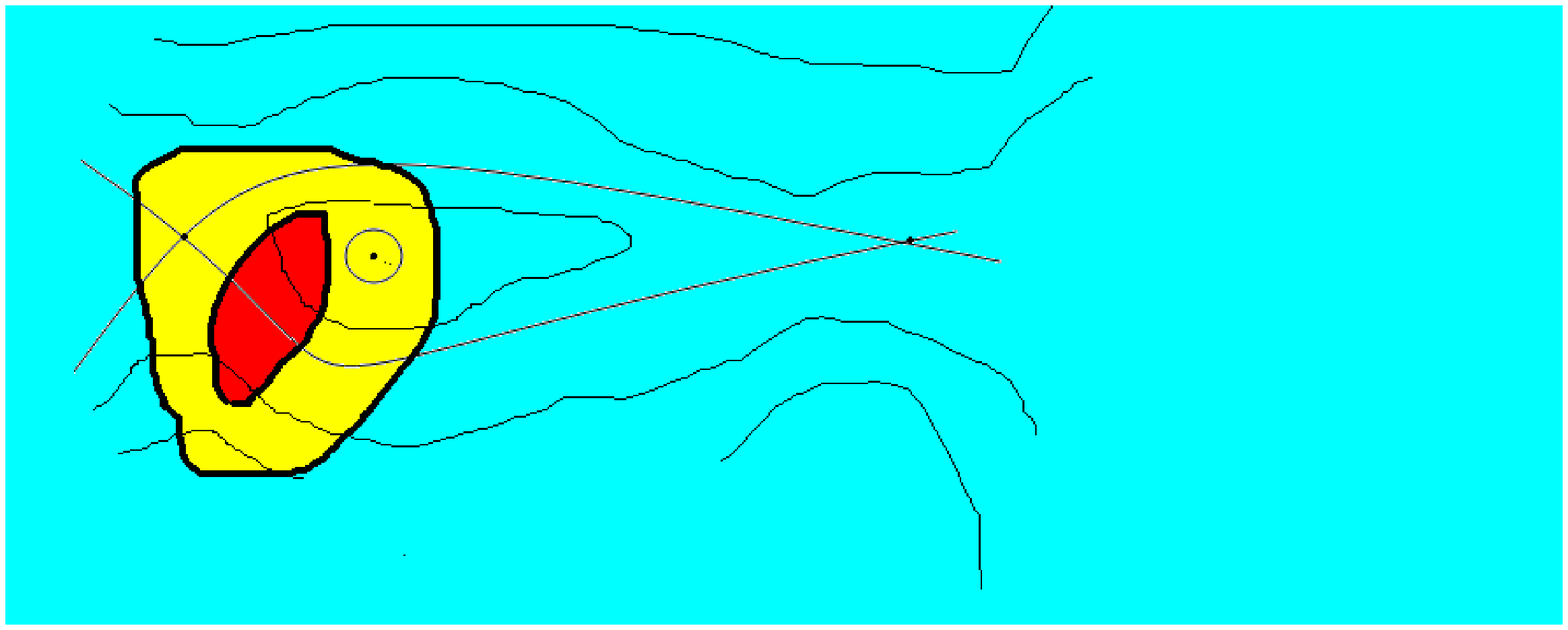}
  \caption{\label{f5}{\bf Symbolic Phase Space for Universe, Level curves for ReH and Color for
    Im H}
  Imaginary part Im H symbolized by colors: Red very strongly wished;
  Yellow also very good, but not the perfect; turquoise strongly to be avoided,
  bad!}
\end{figure}

  \begin{figure}
    \includegraphics[scale=0.7]{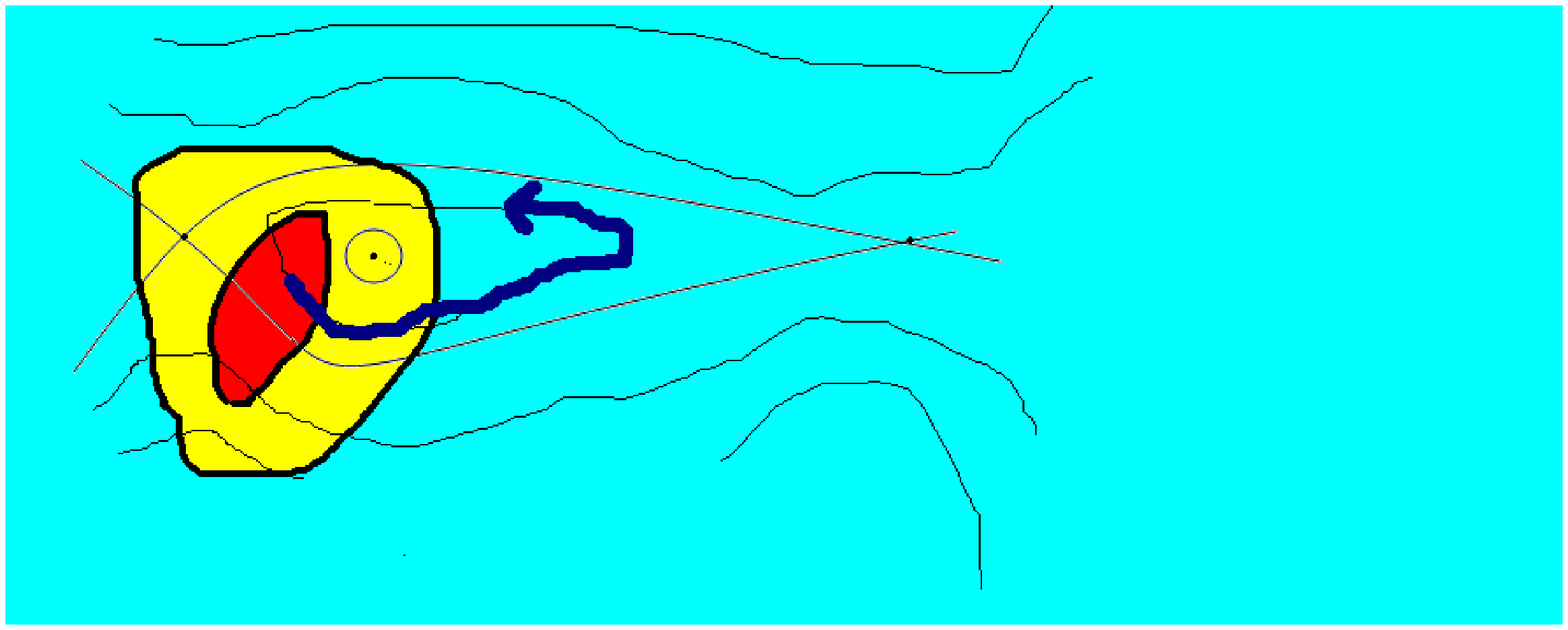}
    \caption{\label{f6}{\bf Symbolic Phase Space for Universe, Level
        curves for ReH
        and Color for
    Im H}
  Imaginary part Im H symbolized by colors: Red very strongly wished;
  Yellow also very good, but not the perfect; turquoise strongly to be avoided,
  bad!}
    \end{figure}

\begin{figure}
  \includegraphics[scale=0.7]{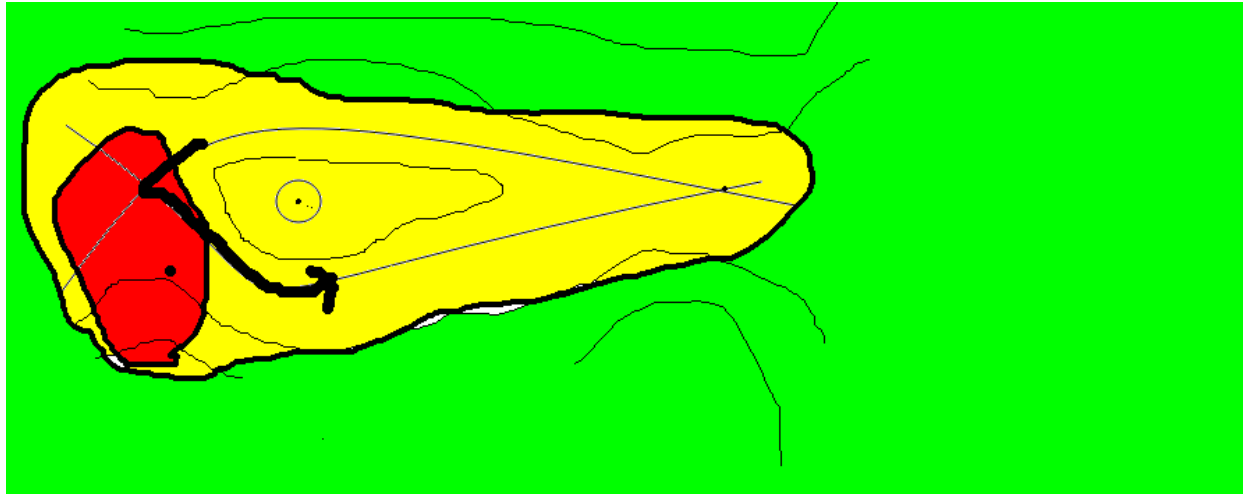}
  \caption{\label{f7}
  {\bf Symbolic Phase Space for Universe, Level curves for Re H and Color for
    Im H}
   Red very strongly wished, the saddle point very good;
   Yellow also very good, but not the perfect, but on this figure
   it can circle around more stably in the yellow; turquoise strongly to be avoided,
   bad, can be avoided by keeping around the fix points!}
\end{figure}

Now some are figures formulated in phase space illustrating these consideration:
see figures \ref{f3},...,\ref{f7}.
How should the system choose to move? To keep red, or yellow, and
    avoid turquoise ?

  \begin{itemize}
  \item It could start in the red to ensure a favorable Im H in the
    start, but alas, it comes out in the turquoise and spend a lot of time with
    very unfavorable Im H.
  \item It could choose a not too bad, i.e. e.g yellow, place with high
    stability so that it can stay there forever and enjoy at least the
    yellow!
  \end{itemize}

\noindent
What we would like to learn from path favorable for high Im H integrated over time?
  What we want to learn from this consideration: It will usually be favorable
  with regard to Im H to choose a very stable place to avoid running around and
  loosing enormously (the turquoise) with regard to Im H.
  So this kind of theory predicts:
  \begin{itemize}
  \item Preferably Universe should be just around a very stable, locally
    ground state, it is the vacuum, with the bottom in the Hamiltonian.
  \item Even better might be using a saddle point (there are also more
    of them and so it more likely to be best) and then choose it so that
    there is a stable local ground state not so far to spend eternity.
    Such a saddle is the tip of the inflaton effective  potential.
    By choosing just that tip in principle it can stand as long as to
    be disturbed by quantum mechanics.
    \end{itemize}

\noindent
Results of the intuitive treatment of the non-Hermitian Hamiltonian are as follows: 
  \begin{itemize}
  \item The world-system should run around so little as possible to avoid
    the low Im H places (= the unfavorable ones): It should have small
    entropy (contrary to the intuitive cosmology of Paul Framptons in
    the other talk). Rather close to stable point (= ground state).
    So the model predicts there being a bottom in the effective Hamiltonian
    locally in phase space. 

  \item The world-system should stay as long as possible at a saddle
    almost exactly - to keep staying surprisingly long (like a pen
    standing vertically on its tip in years); but this is in the many
    degrees of freedom translation a {\bf surprisingly long inflation}
    era. That is the problem of the too many e-foldings, which we thus
    at least claimed to have a feature of the initial condition model helping
    in the right direction (making inflation longer than the potential
    that should preferably not be flat indicates).
    \end{itemize}

\section{Main Result}
\label{main}

Our main result is that you would not discover from equations of motion that 
the Hamiltonian had an anti-Hermitian part. 
  The system (the world) would only have significant Hilbert vector
  components in the states with the very highest imaginary part of the
  eigenvalues of the non-Hermitian Hamiltonian, since the rest would die out
  with time. Remember
 \begin{eqnarray}
    |A(t)> &=& \exp(-i H t)|A(0)>.
  \end{eqnarray}
  So at least the anti-Hermitian part is near to its (supposed) maximum, and
  thus at least less significant.
  We introduce a new inner product making the Hamiltonian H
  normal, i.e. making the Hermitian and the anti-Hermitian parts commute.
So it is unnecessary to assume that the Hamiltonian is Hermitian!
It will show up so in practice anyway!

\section{Weak Value}
\label{WV}

  As a result of our thinking of how to interpret the complex action
  theory we came to the concept already studied by Aharonov et.al.\cite{18,19}, the weak value:
  \begin{eqnarray}
       O_{wv}(t) &=& \frac{<B(t)|O|A(t)>}{<B(t)|A(t)>} \label{weak_value}
  \end{eqnarray}
  or better with time development included:
  \begin{equation}
  O_{wv}(t) = \frac{<B(T_B)|\exp(-i(T_B-t)H) O \exp(-i(t-T_A)H)|A(T_A)>}{
    <B(T_b)|\exp(-i(T_B-T_A)H)|A(T_A)>}, 
  \end{equation}
where we have assumed that the states $|B(t)>$ and $|A(t)>$ time-develop 
according to the following Schr\"{o}dinger equations: 
  \begin {eqnarray}
  \frac{d}{dt}|B(t)> &=& -iH^{\dagger}|B(t)>,\\
  \frac{d}{dt}|A(t)> &=& -iH|A(t)>.
  \end{eqnarray}
Our idea is to use the weak value instead of the usual operator average:  
  \begin{eqnarray}
    O_{av}(t) &=& \frac{<A(t)|O|A(t)>}{<A(t)|A(t)>}. 
  \end{eqnarray}

One motivation is that it may give more natural interpretation of 
functional integrals. 
Usually the answer to how to use functional integrals is: 
  ``You can use it to calculate a time development operator - e.g an S-matrix -
  and then use that to propagate the quantum system in the usual Hilbert space
  formalism.'' 
The weak value has a more beautiful functional integral expression: 
\begin{eqnarray}
O_{wv}(t)  &=& \frac{\int  O \psi_B^* \psi_A \exp(\frac{i}{\hbar} S[path]) {\cal D}path }
    {\int \psi_B^* \psi_A \exp((\frac{i}{\hbar} S[path] ){\cal D}path}, 
\end{eqnarray}
where in the numerator the operator $O$ was inserted at the appropriate time, 
than the usual operator average.

  The usual average and the weak value look a priori  quite different, but
  with what we call the maximization principle, that the absolute value of the denominator of 
Eq.(\ref{weak_value}):  
\begin{equation}
      |<B(t)|A(t)>|=|<B(T_B)|\exp(-i(T_B-T_A)H)|A(T_A)>|
\end{equation}
be maximal for fixed normalization of the two states,
you may see that (at least for Hermitian Hamiltonian) one gets
\begin{eqnarray}
  |B(t)> &\propto & |A(t)>.
\end{eqnarray}

In Ref.\cite{13} 
we have found that one can construct such an inner product $|_Q$ that, 
even if at first the Hamiltonian $H$ is not normal, i.e. if
  \begin{eqnarray}
    [H, H^{\dagger}] &\ne & 0,
  \end{eqnarray}
  then, with regard to this new inner product, it is
  \begin{eqnarray}
    [H, H^{\dagger_Q}] &=&0.
  \end{eqnarray}
The new inner product\footnote{Similar inner products are also studied 
in Refs.\cite{24,25,29}.} can arrange a normal Hamiltonian.

  The inner product can be described as composed from the usual one $|$
  and a Hermitian operator $Q$ constructed from $H$. I.e. $|_Q = | Q$ means 
  \begin{eqnarray}
<...|_Q \; ...> &=& <...|Q|...>.
  \end{eqnarray}
    One can thus talk about a Q-Hermitian operator $O$ when it obeys
    \begin{eqnarray}
      O^{\dagger_Q} =O
\end{eqnarray}
where
\begin{eqnarray}
       O^{\dagger_Q} &=&Q^{-1}O^{\dagger}Q. 
\end{eqnarray}
Remember the point of our new inner product was to make the at
    first not even normal Hamiltonian at least normal, i.e. the
    Q-Hermitian and the anti-$Q$-Hermitian parts commute.
    It is the idea that the physical observables one should use in
    a world with a non-Hermitian Hamiltonian are Q-Hermitian.

\vspace{1cm}

In Ref.\cite{27} we proposed the following theorem ``maximization principle in the future-included complex action theory'':

\vspace{0.5cm}

{ \em As a prerequisite, assume that a given Hamiltonian $H$ is non-normal but
diagonalizable and that the imaginary parts of the eigenvalues of $H$ are
bounded from above, and define a modified inner product $|_Q$ by means
of a Hermitian operator $Q$ arranged so that $H$ becomes normal with respect to $|_Q$. 
Let the two states $|A(t)>$ and $|B(t)>$ time-develop according to 
the Schr\"{o}dinger equations with $H$ and $H^{\dag_Q}$ respectively:
\begin{eqnarray}
|A(t)> &=&
\exp(-iH(t-T_A))|A(T_A)>, \\
|B(t)> &=& \exp(- i H^{\dagger_Q} (t-T_B ))|B(T_B )>, 
\end{eqnarray}
and be normalized with
$|_Q$ at the initial time $T_A$ and the final time $T_B$ respectively:
\begin{eqnarray}
    <A(T_A)|_Q A(T_A)>& =&    1,\\
    <B(T_B )|_Q B(T_B )>& =& 1.
\end{eqnarray}
Next determine $|A(T_A)>$ and $|B(T_B )>$ so as to
maximize the absolute value of the transition amplitude $|<B(t)|_Q A(t)>| =
|<B(T_B )|_Q \exp(-i H(T_B - T_A))|A(T_A)>|$. Then, provided that an operator
$O$ is Q-Hermitian, i.e., Hermitian with respect to the inner product $|_Q$,
i.e. $O^{\dag_Q}= O$, the normalized matrix element of the operator $O$ defined by
\begin{eqnarray}
< O>^{BA}_Q
 &=&\frac{ <B(t)|_Q O|A(t)>}{
  <B(t)|_Q A(t)>}
\end{eqnarray}
becomes real and time-develops under a Q-Hermitian
Hamiltonian.}

\vspace{0.5cm}

We note that this theorem shows that the complex action theory could 
make predictions about initial conditions.

\section{Conclusion}

We have put forward our works of looking at a complex action or better a
non-Hermitian Hamiltonian. Since it would not be easily seen that the
Hamiltonian were indeed non-Hermitian after sufficiently long time and
only showing itself up as it were the initial conditions that were
influenced by the anti-Hermitian part, and even this influence looks
promising, we believe that a complex action of non-Hermitian Hamiltonian
model like the one described has indeed a good chance to be the truth.
Our complex action theory would make predictions about initial conditions. 
An intuitive use of non-Hermitian H suggested explanation for: 
Effective bottom in the Hamiltonian; Long Inflation; Low Entropy. 

One should stress that one should consider it a weaker assumption to
assume a non-Hermitian Hamiltonian than a Hermitian one, in as far as
it is an assumption that the anti-Hermitian part is zero, while
assuming the non-Hermitian Hamiltonian is just {\em allowing the
Hamiltonian to be whatever.} It is only because we are taught about
the Hermitian Hamiltonian from the tradition that we tend to consider
it a new and strange assumption to take the Hamiltonian to be non-Hermitian.

\section*{Acknowledgments}

This work was supported by JSPS KAKENHI Grant Number JP21K03381, and accomplished 
during K.N.'s sabbatical stay in Copenhagen. 
He would like to thank the members and visitors of NBI 
for their kind hospitality and Klara Pavicic for her various kind arrangements and 
consideration during his visits to Copenhagen.  
H.B.N. is grateful to NBI for allowing him to work there as emeritus. 
Furthermore, the authors would like to thank the organizers of Bled workshop 2022 
for their kind hospitality. 



\end{document}